\documentclass[12pt]{article}
\usepackage{amsmath,amsbsy,amssymb,graphics}
\usepackage{graphicx,epsfig}
\def\beq{\begin{equation}}
\def\eeq{\end{equation}}
\def\be{\begin{equation}}
\def\ee{\end{equation}}
\def\bea{\begin{eqnarray}}
\def\eea{\end{eqnarray}}
\def\nnb{\nonumber}

\newcommand{\gsim}{\lower.7ex\hbox{$\;\stackrel{\textstyle>}{\sim}\;$}}
\newcommand{\lsim}{\lower.7ex\hbox{$\;\stackrel{\textstyle<}{\sim}\;$}}

\begin{document}

\begin{center}
 \vspace{0.2cm}
 {\Large \bf Moment equations of neutrinos in supernova  }

\vspace{0.6cm} {\large \bf Wei Liao}

\vspace{0.3cm} {
 Institute of Modern Physics,
 East China University of Science and Technology, \\
 P.O. Box 532, 130 Meilong Road, Shanghai 200237, P. R. China
 \vskip 0.3cm

 Center for High Energy Physics, Peking University, Beijing 100871, P. R. China

}
\end{center}

\begin{abstract}
 \vskip 0.2cm
 We derive a series of moment equations describing the motion and flavor
 transformation of neutrinos in supernova. We find a particular series
 of moments of neutrino density matrix in supernova. The emission angle
 distribution of neutrinos is described by this series of moments. We
 expand the equation of neutrinos using these moments and obtain moment
 equations. We find that these moments have very good property of convergence
 and the infinite series of equations can be truncated to equations with
 a small set of moments. Using a small set of moment equations the required
 computational power is reduced by about two orders of magnitude compared to
 that in multi-angle simulation. The study on non-linear flavor transformation
 of neutrinos is substantially simplified using these equations. Two flavor
 system of neutrinos is also considered and new equations describing the
 flavor polarization vectors of neutrinos are found.

\end{abstract}

PACS: 14.60.Pq, 97.60Bw

\section{Introduction}\label{sec1}
 It is realized that neutrino flavor transformation in supernova is a very
 complicated problem. In addition to the vacuum oscillation effect
 and the effect of neutrino refraction with ordinary matter
 (Mikheyev-Smirnov-Wolfenstein effect), it is
 well known that neutrino-neutrino refraction can also be
 important in supernova. Neutrino self-interaction
 can be important in flavor oscillation when neutrino density is
 sufficiently large. In core collapse supernova neutrino density can be so large
 above the neutrino sphere that the non-linear flavor oscillation caused by
 neutrino-neutrino refraction can dominate neutrino flavor transformation.

 Researches on non-linear neutrino oscillation in supernova have been done by many
 groups. Many interesting phenomena caused by neutrino
 self-interaction, such as synchronized oscillation, bipolar oscillation
 and spectral split, have been found. Detailed
 numerical analysis and qualitative analysis have been done
 to understand these phenomena. An incomplete list of researches on this subject is
 ~\cite{Pan,saw,dfq0,dfcq1,hrsw,dfc,rs1,dfcz,flmm,flmmt,dfcq2,dfq,rs2,dd,
 dfq3,ddmr,ccdk,gv,sf,dfcq4,prs}.
 Although the development in this field is very fast there are
 still many problems not answered and the present situation of
 research is not satisfactory.

 One problem hard to explore is how neutrino flavor transformation depends on
 the emission angle distribution of neutrinos on the neutrino
 sphere. It is known that if all neutrinos are emitted in radial
 direction neutrino self-interaction vanishes and non-linear flavor
 transformation disappears. Angular distribution of neutrino emission
 is essential in non-linear flavor transformation and a complete analysis
 of neutrino flavor transformation has to take it into account.
 However in the present formulation of the problem it is hard to
 study the effect of emission angle distribution of neutrinos.
 Previous researches on this problem use many angle bins
 in numerical simulation. It is very complicated and does not
 give us insight to physics in it.

 In the multi-angle simulation flavor evolutions of neutrinos on trajectories of all
 angle bins should be studied~\cite{dfcq1}. It is found that
 if $L_\nu=10^{51}$ erg$/$s more than 500 angle bins are required in order
 to make the simulation converge. Even more angle bins are
 required for larger neutrino luminosity. Note that the energy range
 is also made discrete in numerical works. It is found that a
 complete numerical computation
 requires following the evolution governed by more than a million
 equations. It is terribly complicated and
 makes us hard to understand the physics in it. Another problem
 in multi-angle simulation is that exact spherical symmetry is essential
 in the simulation. It is unable to imagine how to simulate
 neutrino flavor transformation without the spherical symmetry in
 this approach. A better formulation of the problem is required to
 make it easier to solve numerically and to be understood qualitatively.

  The present scheme to study neutrino evolution is not only
 too complicated in practice but also conceptually not necessary.
 It is not necessary to know the detailed evolution of wave function
 of neutrino in every trajectory. We just need the flux,
 flavor content, angular distribution and energy distribution
 of neutrinos at given time and given position in space. It is enough
 using these information to study the flavor conversion rates of
 neutrinos as functions of radius $r$. It is also sufficient using
 these information to compute the energy deposit of neutrinos
 to plasma environment. Effect of the emission angle distribution of
 neutrinos can be carefully taken into account in
 moment expansion of neutrino distribution in phase space. The whole set of equations
 can be tremendously reduced if appropriate set of moments are found.

 In this article we present a new formulation
 describing the transport and flavor evolution of neutrinos above the neutrino sphere in
 core-collapse supernova. Instead of considering the neutrino flavor
 evolution on trajectories we study the density matrix
 of neutrino at given time and position. We do a moment expansion
 of the Liouville equation of neutrino using a series of moments
 of density matrix.
 The problem of neutrino flavor transformation in supernova is
 tremendously simplified by observing that this series of
 moments have very good property of convergence. Numerical computation
 on neutrino flavor transformation is greatly simplified using
 this series of moment equations. This series of moment equations is
 obtained when we examine the problem using spherical coordinate and
 find a series of functions which have very good convergence property.
 It is shown that the emission angle distribution of
 neutrinos can be systematically studied.

 In section \ref{sec2} we rewrite the Liouville equation
 using spherical coordinate. In section \ref{sec3}
 we simplify the equation of motion with the assumption of spherical
 symmetry. In section \ref{sec4}
 we introduce a series of moments of neutrino density matrix and
 motivate the use of them by examining their property of convergence.
 In section \ref{sec5} we expand the Liouville equation of neutrino
 and work out a series of moment equations.
 In section \ref{sec6} we discuss stationary approximation and
 further simplify the equations. In section \ref{sec7} we discuss the
 truncation of the moment equations.
 In section \ref{sec8} we present some numerical analysis.
 In section \ref{sec9} we work in two flavor system and derive equations
 governing the pendulum motion of neutrinos in flavor space.
 We conclude in section \ref{sec10}.

\section{Equation in spherical coordinate } \label{sec2}
 In this section we derive the equation of motion of neutrinos in
 spherical coordinate. The neutron star core of the core-collapse
 supernova is located at the origin of the coordinate system.

     \begin{figure}
\begin{center}
\includegraphics[height=5.cm,width=7.8cm]{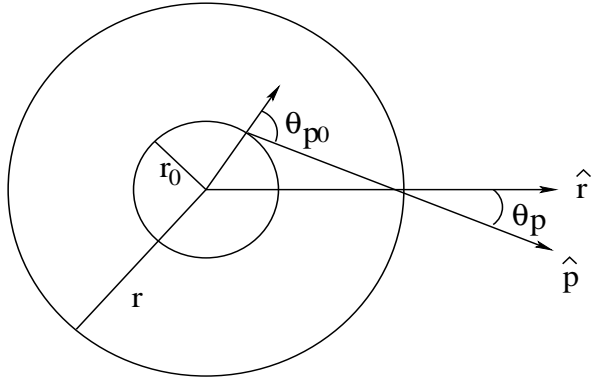}
\end{center}
 \vskip 0.0cm
  \caption{\small Geometric picture of angles of the neutrino momentum
  intersect with ${\hat r}$.}
 \label{fig1}
\end{figure}

 Consider the density matrix $\rho_{\vec p}(t,{\vec x})$ of neutrino with a given
 momentum ${\vec p}$ at position ${\vec x}$ and at time $t$.
 $\rho_{\vec p}(t,{\vec x})$ is a $3\times 3$ matrix
 when considering three flavors of neutrinos and is a
 $2\times 2$ matrix when considering two flavors of neutrinos.
 For anti-neutrino we express the density
 matrix as ${\bar \rho}_{\vec p}$. The evolution of neutrino
 flavor is described by the Liouville equation~\cite{sr,mt}:
 \bea
 \frac{d \rho_{\vec p}}{d t}=-i [H_0+\sqrt{2} G_F L+\sqrt{2}
 G_F D , \rho_{\vec p}], \label{LVEq1}
 \eea
 where $D$ describes neutrino self-interaction and is expressed as
 \bea
 D=\int \frac{d^3 q}{(2 \pi)^3} (\rho_{\vec q}-{\bar \rho}_{\vec q})
 (1-\cos\theta_{{\vec p} {\vec q}}), \label{Dterm}
 \eea
 and $H_0$ is the Hamiltonian for vacuum oscillation,
 $L=diag\{n_e,n_{\mu},n_{\tau}\}$ in the flavor base is the matter
 term given by charged lepton number densities $n_{e,\mu,\tau}$.
 $G_F$ is the Fermi constant.
 The right-hand side of Eq. (\ref{LVEq1}) is a commutator of the
 effective Hamiltonian and the density matrix. For anti-neutrino
 the equation is similar except replacing $H_0$ by $-H_0$:$H_0 \to - H_0$.

 The right-handed side of Eq. (\ref{LVEq1}) governs the quantum evolution of
 neutrino flavor. It does not include the effect of streaming of
 relativistic neutrinos. Effect of the motion of neutrinos is
 implicitly given in the left-hand side of Eq. (\ref{LVEq1})
 which can be written as
 \bea
 \frac{d \rho_{\vec p}}{d t}=\frac{\partial \rho_{\vec p}}{\partial t}
 +\frac{d x^i}{d t} \frac{\partial \rho_{\vec p}}{d x^i}+
 \frac{d p^i}{d t} \frac{\partial \rho_{\vec p}}{d p^i}. \label{LVEq1b}
 \eea
 The second term in Eq. (\ref{LVEq1b}) is caused by the streaming of
 neutrino in space and the third term is caused by the time
 dependence of neutrino momentum.  For free streaming neutrino
 the third term is zero in Cartesian coordinate. It does not
 vanish in spherical coordinate: ${\vec x}=(r,\theta,\varphi)$.
 Fig. \ref{fig1} demonstrates that $\theta_p$, the
 angle of neutrino direction intersecting with
 the radial direction, changes as neutrino propagates.
 To figure out the left-hand side of Eq. (\ref{LVEq1}) we need
 to know $\frac{d x^i}{d t}$ and $\frac{d p^i} {d t}$.

\begin{figure}
\begin{center}
\includegraphics[height=6.cm,width=6cm]{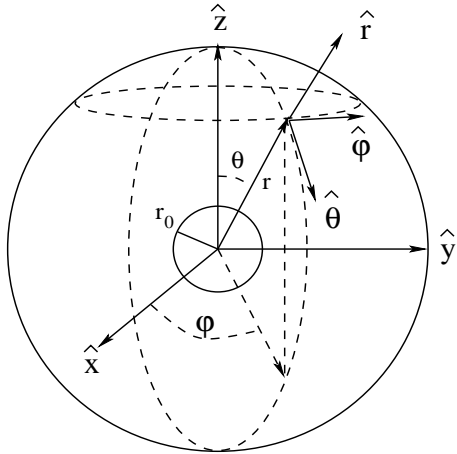}
\end{center}
 \vskip 0.0cm
  \caption{\small Spherical coordinate
  and the local coordinate at ${\vec x}=(r,\theta,\varphi)$.}
 \label{fig2}
\end{figure}

 The spherical coordinate system is shown in Fig. \ref{fig2}.
 At position ${\vec x}$ we introduce a local coordinate system spanned
 by three orthogonal unit vectors: ${\hat r}$, ${\hat \theta}$
 and ${\hat \varphi}$:
 \bea
 &&{\hat r}=\sin\theta \cos\varphi {\hat x}+\sin\theta \sin\varphi
 {\hat y}+ \cos\theta {\hat z}, \label{locCoord1}\\
 &&{\hat \theta}=\cos\theta \cos\varphi {\hat x}+\cos\theta \sin\varphi
 {\hat y}-\sin\theta {\hat z}, \label{locCoord2}\\
 && {\hat \varphi}=-\sin\varphi {\hat x}+ \cos\varphi {\hat y},
 \label{locCoord3}
 \eea
 where ${\hat x}$, ${\hat y}$ and ${\hat z}$ are three orthogonal unit
 vectors in Cartesian coordinate. They are shown in Fig. \ref{fig2}.
 The momentum ${\vec p}$ of a
 neutrino at position ${\vec x}$ can be projected to ${\hat r}$,
 ${\hat \theta}$ and ${\hat \varphi}$ directions.
 $\theta_p$ and $\varphi_p$ are defined as:
 \bea
 {\hat p} \cdot {\hat r}=\cos\theta_p,
 ~~{\hat p}\cdot {\hat \theta}=\sin\theta_p \cos\varphi_p,
 ~~{\hat p} \cdot {\hat \varphi}= \sin\theta_p \sin\varphi_p, \label{locCoord4}
 \eea
 where ${\hat p}={\vec p}/|{\vec p}|$ is a unit vector.
 In Appendix we compute the projection of momentum to three
 directions. Using Eq. (\ref{momentum}) we get
 \bea
 && \frac{d r}{d t}= \frac{p^r}{p^0} = \cos\theta_p, \label{velocity1} \\
 && \frac{d \theta}{d t}= \frac{p^\theta}{p^0}=\frac{1}{r}
  \sin\theta_p \cos\varphi_p, \label{velocity2} \\
 && \frac{d \varphi}{d t}= \frac{p^\varphi}{p^0}=\frac{1}{r \sin\theta}
 \sin\theta_p \sin\varphi_p, \label{velocity3}
 \eea
where $p^0$ is the zero component of momentum.

 $\frac{d p^i}{d t} $ is given by the geodesic equation:
 \bea
 \frac{d p^i}{dt} = - \Gamma^i_{jk} ~p^j p^k/p^0,\label{GeodEq}
 \eea
 where $\Gamma^{i}_{jk}$ is the Christoffel symbol and summation over
 repeated indices is assumed. Equations of
 $p^r$, $p^{\theta}$ and $p^{\varphi}$ are computed in the Appendix.
 It is shown that $|{\vec p}|$, the magnitude of neutrino momentum,
 is a constant. This is an expected result when gravitational
 potential is neglected. Hence
 in Eq. (\ref{LVEq1b}) we just need to include effect of angular
 dependence of neutrinos. Writing
 $\rho_{\vec p}(t,{\vec x},{\vec p})
 =\rho_{\vec p}(t, {\vec x},|{\vec p}|, \cos\theta_p,\varphi_p)$
 (similarly for ${\bar \rho}_{\vec p}$) we just need to include
 $d \cos\theta_p/dt$ and $d \varphi_p/dt$ in Eq. (\ref{LVEq1b}). They are given
 in Eqs. (\ref{GeodEq5}) and (\ref{GeodEq6}) in Appendix. To summarize
 we arrive at
 \bea
\frac{d \rho_{\vec p}}{d t}&& =\frac{\partial \rho_{\vec
p}}{\partial t} + \cos\theta_p \frac{\partial \rho_{\vec
p}}{\partial r} +\frac{1}{r} \sin\theta_p \cos\varphi_p
\frac{\partial \rho_{\vec p}}{\partial \theta} + \frac{1}{r
\sin\theta} \sin\theta_p \sin \varphi_p\frac{\partial \rho_{\vec
p}}{\partial \varphi} \nnb \\
&& + \frac{1}{r}(1-\cos^2\theta_p)\frac{\partial \rho_{\vec
p}}{\partial \cos\theta_p}-\frac{1}{r} ctg\theta \sin\theta_p
\sin\varphi_p \frac{\partial \rho_{\vec p}}{\partial \varphi_p}.
\label{LVEq2a}
 \eea

 \section{Equation with spherical symmetry}\label{sec3}

 In the remaining part of the present article we assume
 that the system in core-collapse supernova is
 spherically symmetric and the angular distribution of
 neutrino momentum at position ${\vec x}=(r,\theta,\varphi)$ is
 cylindrically symmetric, that is
 \bea
 && \frac{\partial L}{\partial \theta}=\frac{\partial L}{\partial
 \varphi}=0,~~ \label{assump0} \\
 && \frac{\partial \rho_{\vec p}}{\partial \theta}=
 \frac{\partial \rho_{\vec p}}{\partial \varphi}=0,~~
 \frac{\partial \rho_{\vec p}}{\partial \varphi_p}=0. \label{assump}
 \eea
  We obtain
 \bea
 \frac{\partial \rho_{\vec p}}{\partial t} + \cos\theta_p
 \frac{\partial \rho_{\vec p}}{\partial r}
 + \frac{1}{r}(1-\cos^2\theta_p)\frac{\partial \rho_{\vec
 p}}{\partial \cos\theta_p}=-i [H_0+\sqrt{2} G_F (L+ D), \rho_{\vec
 p}].
 \label{LVEq2}
 \eea
 Alternatively we introduce
 \bea
 \mu_p=1-\cos\theta_p. \label{mup}
 \eea
 Eq. (\ref{LVEq2}) becomes
 \bea
 \frac{\partial \rho_{\vec p}}{\partial t} + (1-\mu_p)
 \frac{\partial \rho_{\vec p}}{\partial r}
 - \frac{1}{r}\mu_p (2-\mu_p) \frac{\partial \rho_{\vec
 p}}{\partial \mu_p}=-i [H_0+\sqrt{2} G_F (L+ D), \rho_{\vec
 p}].
 \label{LVEq2b}
 \eea
 $\rho_{\vec p}=\rho_{\vec p}(t,r,|{\vec p}|,\mu_p)$ is the density matrix
 at time $t$ and radius $r$. $|{\vec p}|$ and $\mu_p$ are variables used
 for describing distribution of neutrinos in momentum space. $\mu_p$ takes value
 in the range $[0,2]$.

 We note that for a realistic distribution of supernova neutrinos the density
 matrix $\rho_{\vec p}$ should be a smooth function of $|{\vec p}|$
 and $\mu_p$. Moreover, $\rho_{\vec p}$ should decrease to zero
 as $\mu_p \to 1$, i.e. as $\theta_p \to \pi/2$. This is because
 no neutrinos inside the neutrino sphere can be emitted in the
 direction with $\theta_{p0} \ge \pi/2$ on the neutrino sphere
 where $\theta_{p0}$ is the emission angle on the neutrino sphere
 as shown in Fig. \ref{fig1}.
 \footnote{$\theta_{p0}=\pi/2$ at $r=r_0$ means that it is a trajectory
 passing through the surface of the sphere. No neutrino can be emitted in
 this direction on the neutrino sphere.}
 Neutrinos moving out of the supernova have $\theta_p < \pi/2$ for $r \ge r_0$.
 This is expressed by the
 following condition for the density matrix
 \bea
 \rho_{\vec p}(t,r,|{\vec p}|,\mu_p)=0 , ~~\textrm{for} ~\mu_p \geq 1
 ~\textrm{and} ~r\ge r_0. \label{conddirect}
 \eea
 This condition will be used when we do moment expansion of equation.

 A simple application of Eq. (\ref{LVEq2}) is for the total
 intensity of neutrino
 \bea
 f_{\vec p} = Tr[\rho_{\vec p}], \label{intensity}
 \eea
 where $Tr$ is the trace of matrix.
 The right-hand side of Eq. (\ref{assump}) is a commutator of matrices
 and its trace is zero. We find
 \bea
\frac{\partial f_{\vec p}}{\partial t} + \cos\theta_p \frac{\partial
f_{\vec p}}{\partial r} +
\frac{1}{r}(1-\cos^2\theta_p)\frac{\partial f_{\vec p}}{\partial
\cos\theta_p}=0. \label{Boltz}
 \eea
 We integrate Eq. (\ref{Boltz}) by $\int d\Omega_{\vec p}$ which is over all
 solid angle.  Since the integration limit does not depend on $r$, we can
 change the order of $\frac{\partial}{\partial r}$ and the integration symbol
 $\int d\Omega_{\vec p}$.
 \footnote{More discussion on this point is given in section \ref{sec5}.}
 After integration by part using condition Eq. (\ref{conddirect})
 we find
 \bea
 \frac{\partial f^0_p}{\partial t}
 +\frac{\partial (f^0_p \beta)}{\partial r} +\frac{2}{r} f^0_p \beta=0,
 \label{diffusion0}
 \eea
 where
 \bea
 f^0_p=\int d\Omega_{\vec p} ~f_{\vec p},
 ~f^0_p \beta=\int d\Omega_{\vec p} ~\cos\theta_p~f_{\vec p}.
 \label{diffusion1}
 \eea

 The simplest case is that neutrinos are all emitted in radial direction
 at neutrino sphere. Hence $\beta=1$ and Eq. (\ref{diffusion0}) is
 simplified as
 \bea
\frac{\partial f^0_p}{\partial t}
 +\frac{1}{r^2} \frac{\partial (r ^2 f^0_p)}{\partial r}=0.\label{diffusion2}
 \eea
 The solution of this equation is
 \bea
 f^0_p(t,r)=\frac{r_0^2}{r^2} f^0_p(t-(r-r_0),r_0). \label{diffusion}
 \eea
 $r_0$ is the radius of neutrino sphere.
 The meaning of this solution is obvious. Intensity of neutrino
 at time $t$ and radius $r$ is directly related to the
 neutrino intensity on neutrino sphere at time $t-(r-r_0)$.
 Intensity is suppressed by the geometric factor
 $\frac{r_0^2}{r^2}$, as expected in a spherically symmetric system.

 An interesting observation based on Eq. (\ref{diffusion}) is that
 if emission of neutrino on neutrino sphere (at $r=r_0$) has very weak dependence
 on time, $\frac{\partial f^0}{\partial t}=0$ is obtained for all
 $r> r_0$. It means Eq. (\ref{diffusion}) gives a stationary
 distribution of neutrino density in space. We denote the
 approximation $\frac{\partial \rho}{\partial t}=0$ as the
 stationary approximation. It is very useful for simplifying the
 moment equations. We come back to this approximation later.

\section{Moments of density matrix}\label{sec4}
 In this section we examine in detail the geometric characteristics
 of neutrino transport in supernova. We introduce moments
 of density matrix and check their convergence properties.
 We motivate moment expansion of Eq. (\ref{LVEq2b}).

 Consider a neutrino emitted at neutrino sphere at position
 ${\vec x}_0=(r_0,\theta_0,\varphi_0)$
 with an emission angle $\theta_{p0}$ intersecting with the radial direction.
 It is shown in Fig. \ref{fig1}. At
 time $t$ neutrino arrives at position ${\vec x}=(r,\theta,\varphi)$
 with an angle $\theta_p$ intersecting with the radial direction ${\hat r}$.
 From Fig. \ref{fig1} one can see that
 \bea
 \sin\theta_p=\frac{r_0}{r} \sin\theta_{p0}. \label{geom1}
 \eea
 From this we get
 \bea
 \cos\theta_p = \sqrt{1-\frac{r^2_0}{r^2} \sin^2 \theta_{p0} }
 \label{geom2}
 \eea
 Neutrinos are more peaked in radial direction if $r$ is larger.
 We can see that at radius $r$ the condition Eq. (\ref{conddirect})
 says that
 \bea
 \rho_{\vec p}=0, ~\textrm{for}~ \sin\theta_p \geq \frac{r_0}{r}
 ~\textrm{or} ~\cos\theta_p \leq \sqrt{1-\frac{r^2_0}{r^2}}.
 \label{geom3}
 \eea
 Alternatively Eq. (\ref{geom3}) is re-expressed as
 \bea
 \rho_{\vec p}=0,~\textrm{for}~\mu_p \geq  S(r) ~\textrm{at radius} ~r,\label{geom5}
 \eea
 where $S(r)$ is the geometric scaling factor:
 \bea
 S(r)=1-\sqrt{1-r^2_0/r^2}=
 \frac{r^2_0/r^2}{1+\sqrt{1-r^2_0/r^2}}.
 \label{geom6}
 \eea
 $S(r)$ scales as $r^{-2}$.

 We introduce $k$th moment of the density matrix:
 \bea
 \rho_k(t,r,p=|{\vec p}|)=\int d\Omega_{\vec p} ~ \mu_p^k ~\rho_{\vec p}(t,r),
 ~k=0,1,2\cdots. \label{rhok}
 \eea
 The integration is over all solid angle of neutrinos.
 It's easy to see that $\rho_k$ converge very fast as neutrinos propagate
 out of supernova. Note that $\int_{-1}^1 d\cos\theta_p=\int_0^2 d \mu_p$ and
 using (\ref{geom5}) we find
 \bea
 \rho_k(t,r,p) =\int_0^2 d\mu_p \int_0^{2\pi} d\varphi_p~\mu_p^k~\rho_{\vec p}
 =\int_0^{S(r)} d\mu_p \int_0^{2\pi} d\varphi_p~\mu_p^k~\rho_{\vec p}.\label{geom7}
 \eea
 Because $\rho_{\vec p}=0$ for $\mu_p \geq S(r)$ the integration over $\mu_p$ is
 effectively restricted to the range $[0,S]$.
  $\mu_p^k$ factor in the integration gives a strong suppression.
 The larger the number k is, the stronger the suppression is. From Eq. (\ref{geom7})
 we note that $\rho_k$ is suppressed roughly by $S^{k+1}$:
 \bea
 \rho_k \propto S^{k+1}. \label{geom8}
 \eea

 $\rho_0$ scales as $S(r)$, i.e. approximately $r^{-2}$, if neglecting effect of
 flavor conversion. This is the geometric factor shown in
 Eq. (\ref{diffusion}) for the total intensity.
 Comparing with $\rho_0$ $\rho_k$ is further suppressed by the factor
 $\mu_p^k$ in the integration and scales as $S^{k+1}(r)$, i.e.
 approximately $r^{-2(k+1)}$. Hence at large $r$ higher moments can be
 safely neglected. It's clear that this set of moments
 has very good convergence property. If we use this set of moments
 to expand  Eq. (\ref{LVEq2b}) we should have a series of equations
 which has very good convergence property.

 Moments given in Eq. (\ref{rhok}) are very nice
 quantities supernova offered to us in studying flavor
 conversion of neutrinos. In section \ref{sec5}
 we will use this series of moments to expand Eq. (\ref{LVEq2b}).

\section{Moment expansion of the equation of motion}\label{sec5}
 In this section we derive moment equations describing the transport
 and flavor transformation of neutrinos in supernova.

 First, we examine the right-hand side of Eq. (\ref{LVEq2b})
 and express $D$, the self-interaction term, in terms of $\rho_k$.
 Notice that Eq. (\ref{assump}) says
 \bea
 \int d^3 q ~ \rho_{\vec q} ~(1-\cos_{{\vec p} {\vec q}})
 =\int d^3 q ~ \rho_{\vec q} ~ (1-\cos\theta_q\cos\theta_p),
 \label{expand1}
 \eea
 where $\theta_q$ and $\theta_p$ are angles of ${\vec q}$ and
 ${\vec p}$ separately. Using
 \bea
 1-\cos\theta_q\cos\theta_p=(1-\cos\theta_q)+(1-\cos\theta_p)
 -(1-\cos\theta_q)(1-\cos\theta_p), \label{expand2}
 \eea
 we can get
 \bea
 D=(n_0-{\bar n}_0) \mu_p+ (n_1-{\bar n}_1) (1-\mu_p),
 \label{expand3}
 \eea
 where
 \bea
  n_0(t,r)&&=\int \frac{d^3 q} {(2 \pi)^3} ~\rho_{\vec q}
 = \int \frac{d|{\vec q}|}{(2 \pi)^3} ~|{\vec q}|^2 ~\rho_0(t,r,|{\vec q}|). \label{expand4} \\
  n_1(t,r)&& =\int \frac{d^3 q} {(2 \pi)^3} ~(1-\cos\theta_q) \rho_{\vec q}
  =\int \frac{d{|\vec q}|}{(2 \pi)^3} ~|{\vec q}|^2 ~\rho_1(t,r,|{\vec q}|).
  \label{expand5}
 \eea
 Similarly for ${\bar n}$ and ${\bar \rho}$.

 We integrate Eq. (\ref{LVEq2b}) using $\int d\Omega_{\vec p} ~\mu_p^k
 = \int^{2\pi}_0 d\varphi_p \int^2_0 d\mu_p ~\mu_p^k$ which is over all
 solid angle of neutrinos. Since the integration limit does not depend on $r$
 we find that
 $\int d\Omega_{\vec p} ~\mu_p^k ~\frac{\partial}{\partial r}\rho_{\vec p}
 =\frac{\partial}{\partial r}(\int d\Omega_{\vec p} ~\mu_p^k ~\rho_{\vec
 p})$.
 \footnote{ Since $\rho_{\vec p}$ vanishes in some region effectively there is an
  integration limit which depends on $r$ as shown in Eq. (\ref{geom7}).
  This effective limit does not change the fact that
  the differential and integral operators can be interchanged. It
  can be checked as follows.
 $\frac{\partial}{\partial r}(\int d\mu_p ~\mu_p^k ~\rho_{\vec
 p})=\frac{\partial}{\partial r}(\int_0^{S(r)} d\mu_p ~\mu_p^k ~\rho_{\vec
 p})=\int d\mu_p ~\mu_p^k ~\frac{\partial}{\partial r}\rho_{\vec
 p}+ (\frac{dS(r)}{dr} \mu_p^k ~\rho_{\vec p})|_{\mu_p=S(r)}=
 \int d\mu_p ~\mu_p^k ~\frac{\partial}{\partial r}\rho_{\vec
 p}$. In the last step $\rho_{\vec p}(r,\mu_p=S(r))=0$ from
 Eq. (\ref{geom5}) has been used.}
 We do integration by part using the
 condition Eq. (\ref{conddirect}) which is universal for all $r\ge r_0$.
 We get
 \bea
 &&\frac{\partial \rho_k}{\partial t}+\frac{\partial (\rho_k-\rho_{k+1})
 }{\partial r}+\frac{1}{r} [2(k+1) \rho_k-(k+2) \rho_{k+1}]
  =-i [H_A, \rho_k]-i[H_B,\rho_{k+1}],\label{LVEq3}
 \eea
 where
 \bea
 H_A=H_0+\sqrt{2}G_F(L+D_1),~~H_B=\sqrt{2} G_F D_0, \label{expand6}
 \eea
 and
 \bea
 D_0=n_0-{\bar n}_0-(n_1-{\bar n}_1),~~D_1=n_1-{\bar n}_1.
 \label{expand6b}
 \eea
 In the remaining part of the article we will also use
 \bea
 H_C=H_A+H_B \label{expand6c}
 \eea

 Eq. (\ref{LVEq3}) can be rewritten as
 \bea
 \frac{\partial \rho_k}{\partial t}
 +\frac{1}{r^{2(k+1)}} \frac{\partial}{\partial r}(r^{2(k+1)} \rho_k)
 -\frac{1}{r^{k+2}} \frac{\partial}{\partial r}(r^{k+2} \rho_{k+1})
 =-i[H_A,\rho_k]-i[H_B,\rho_{k+1}]. \label{LVEq3b}
 \eea

 For $k=0$ and $k=1$ Eq. (\ref{LVEq3b}) gives
 \bea
 &&\frac{\partial \rho_0}{\partial t}+\frac{1}{r^2} \frac{\partial (r^2
 \rho_0) }{\partial r}-\frac{1}{r^2}\frac{\partial (r^2 \rho_1)}{\partial r}
  =-i [H_A, \rho_0]-i[H_B,\rho_1],\label{MNExpand1}\\
  &&\frac{\partial \rho_1}{\partial t}+\frac{1}{r^4} \frac{\partial
(r^4 \rho_1) }{\partial r}-\frac{1}{r^3}\frac{\partial (r^3
\rho_2)}{\partial r}
 =-i [H_A, \rho_1]-i[H_B,\rho_2]. \label{MNExpand2}
 \eea
 If neglecting other terms the first two terms in left-hand side of
 Eq. (\ref{MNExpand1}) make $\rho_0$ scales as $r^{-2}$ which is the correct
 geometric factor as shown in Eq. (\ref{geom8}). The presence
 of $\rho_1$ says that the average velocity of neutrino gas
 is smaller than the speed of light. It modifies the scaling
 behavior of $\rho_0$. The first two terms in the left-hand
 side of Eq. (\ref{MNExpand2})
 give the geometric scaling factor $r^{-4}$ as observed
 in Eq. (\ref{geom8}). The presence of $\rho_2$ in the
 left-hand side of Eqs. (\ref{MNExpand2})
 modifies the scaling behavior of $\rho_1$. Similarly,
 the left-hand side of Eq. (\ref{LVEq3}) makes $\rho_k$ scale as
 $r^{-2(k+1)}$ if neglecting $\rho_{k+1}$ and the presence of
 $\rho_{k+1}$ modifies the scaling behavior.

 Eq. (\ref{LVEq3}) systematically takes into account the
 effect of the angular distribution of neutrino emission.
 The effect is encoded in moments
 introduced in the present article. In particular, the scaling
 law of the strength of the effective Hamiltonian is modified when including
 higher moments $\rho_k$ with $k \geq 2$.
 Furthermore, this series of
 moment equations has very good property of convergence
 and is better for studying flavor transformation of neutrino
 in supernova, for studying non-linear flavor transformation
 in particular. This virtue will help us to truncate the
 infinite series moment equations and will be further discussed in
 section \ref{sec7}.

 An important point when studying effect of self-interaction in flavor conversion
 is that the minimal set of equations should
 include equations for $\rho_0$ and $\rho_1$, Eqs. (\ref{MNExpand1})
 and (\ref{MNExpand2}), because
 if neutrinos are all emitted in radial direction, that is
 $\rho_k=0$ for $k>0$, neutrino self-interaction vanishes.

 \section{Stationary approximation} \label{sec6}
 In solving the equations of neutrino evolution in supernova
 we assume the stationary approximation:
 \bea
 \frac{\partial \rho_k}{\partial t}=0. \label{StatApp}
 \eea
 This approximation is assumed by noticing that the
 neutrino self-interaction is
 important in the region of radius less than hundreds
 kilometers in which neutrino density is sufficiently large.
 Neutrino travels through this region in $\sim 10^{-3}$s. This
 time scale is much shorter than the time scale of neutrino emission
 in supernova which is around $10$s. Therefore, we can assume
 that neutrino luminosity and energy do not change in the time scale
 as short as around $\sim 10^{-3}$s.
 Furthermore, we can also assume that the distribution of ordinary
 matter in supernova does not change in such a short time scale.
 This assumption is valid if the disturbance of ordinary matter
 propagates with a speed much smaller than the speed of light.

 An example supporting this assumption is given by examining
 neutrino intensity given in Eq. (\ref{diffusion}). According
 to the formula it is easy to see that if $\frac{\partial f}{\partial t}=0$
 is assumed at the neutrino sphere where $r=r_0$, $\frac{\partial f}{\partial t}=0$
 is satisfied at $r> r_0$.

 We introduce
 \bea
 \rho^\prime_k=z^{2(k+1)} ~\rho_k,
 ~~{\bar \rho}^\prime_k=z^{2(k+1)} ~{\bar \rho}_k\label{Eqset0}
 \eea
 where
 \bea
 z=\frac{r}{r_0}. \label{Eqset0b}
 \eea
 $\rho^\prime_k$ and ${\bar \rho}^\prime_k$ correctly take
  into account the geometric scaling factor.

 With the assumption of stationary approximation for the density matrix
 we get equation
 \bea
 && \frac{d \rho^\prime_k}{dr}=z^k
 \frac{ d}{d r}( z^{-(k+2)} \rho^\prime_{k+1})
  -i [H_A, \rho^\prime_k]-i[H_B,z^{-2} \rho^\prime_{k+1}], ~k=0,1\cdots. \label{LVEq5}
 \eea
 Using Eq. (\ref{LVEq5}) repeatedly we can get
 \bea
 \frac{d \rho^\prime_k}{d r}&& =-i[H_A,\rho^\prime_k]
  +z^{k-n+1} \frac{d}{dr}(z^{-(k+n+1)} \rho^\prime_{k+n})
 -i[H_B,z^{-2n} \rho^\prime_{k+n}] \nnb \\
 &&- r_0^{-1} \sum_{i=1}^{n-1}(k+i+1)z^{-(2i+ 1)}
 \rho^\prime_{k+i}-i[H_C, \sum_{i=1}^{n-1} z^{-2i}
 \rho^\prime_{k+i}], \label{LVEq5b}
 \eea
 where $n\geq 2$. For $n=1$ the second line in Eq. (\ref{LVEq5b})
 disappears and the equation reduces to Eq. (\ref{LVEq5}).
 Equation of ${\bar \rho}'$ can be similarly obtained for anti-neutrinos.

 \section{Truncation of moment equations} \label{sec7}
 In practical computation we have to truncate the infinite series of
 moment equations. If magnitude of higher moments
 are much smaller than $\rho_{0,1}$ we can assume that
 $\rho_k=0$ for $k> N$ where $N$ is an integral. In this approximation we
 get a set of $2(N+1)$ equations for $\rho_k({\bar \rho}_k)$ or
 $\rho^\prime_k({\bar \rho}^\prime_k)$ where
 $k=0,1\cdots, N$.  We denote this approximation as $P_N$
 approximation.
 In $P_N$ approximation we get
 \bea
 \frac{d \rho^\prime_k}{d r}&& =- r_0^{-1} Q^1_k-i[H_A,\rho^\prime_k]
 -i[H_C, Q^2_k], \label{LVEq5c}
 \eea
 where $k=0,1,\cdots,N$ and
 \bea
 Q^1_k= z^{2k} \sum_{l=k+1}^{N}(l+1)z^{-(2l+ 1)} \rho^\prime_l,~~
 Q^2_k= z^{2k} \sum_{l=k+1}^{N} z^{-2l} \rho^\prime_l. \label{LVEq5cb}
 \eea
 $Q^{1,2}_N=0$. Similarly we have equations for ${\bar \rho}^\prime_k$.

 An appropriate choice of truncation depends both on the
 initial condition and on the property of convergence of these
 moments. It is interesting to note that higher moments are
 naturally suppressed at the neutrino sphere. For example,
 \bea
 \rho_k(t, r_0)= \frac{1}{k+1} \rho_0(t, r_0) \label{approx1}
 \eea
 is obtained using Eq. (\ref{rhok}) if neutrino is uniformly emitted with
 respect to the emission angle $\theta_{p0}$.
 \footnote{As noted previously, for realistic distribution of neutrinos
 $\rho_{\vec p}(r_0)$ should decrease to zero as $\theta_{p0} \to \pi/2$.
 Uniform distribution can be taken as an approximation to a
 distribution which is uniform in the range $0\le \mu_{p0} \le
 1-\epsilon$ and decreases to zero rapidly in the range
 $1-\epsilon \le \mu_{p0} \le 1$ where
 $\epsilon \ll 1$ is taken as a small positive number. This is a good
 approximation for not too large $k$. }
 $\rho_{10}$ is about $10$ times smaller than $\rho_0$.
 For a practical distribution which may have sharp falloff
 at an angle close to $\theta_{p0}=\pi/2$($\mu_{p0}=1$), higher
 moments are further suppressed.
 As an example, suppose the distribution is uniform in the
 region $0 \leq \mu_{p0} \leq 0.9$ and drops to zero rapidly
 at $\mu_{p0}=0.9$ we find that
 \bea
 \rho_{10}(t,r_0) && \approx \frac{1}{11} 0.9^{10} ~\rho_0(t, r_0) \approx 3 \% ~\rho_0(t,r_0),
 \label{approx2}\\
 \rho_{10}(t,r_0) && \approx \frac{2}{11} 0.9^9 ~\rho_1(t, r_0) \approx 7 \% ~\rho_1(t,r_0)
  \label{approx3}
 \eea
 $\rho_{10}$ has magnitude smaller than $\rho_0$ and $\rho_1$,
 the major quantities in the evolution problem.

 Other examples of distribution can also be checked.
 Considering distribution
 proportional to $1-\mu_{p0}$ or $\mu_{p0} (1-\mu_{p0})$ which are
 zero at $\mu_{p0}=1$ we find for these two examples
 \bea
 && \rho_k(t,r_0)=\frac{2}{(k+1)(k+2)}~\rho_0(t,r_0), \label{approx4}\\
 && \rho_k(t,r_0)=\frac{6}{(k+1)(k+2)}~\rho_1(t, r_0). \label{approx4b}
 \eea
 or
 \bea
 && \rho_k(t,r_0)=\frac{6}{(k+2)(k+3)}~\rho_0(t,r_0),
 \label{approx5} \\
 && \rho_k(t,r_0)=\frac{12}{(k+2)(k+3)}~\rho_1(t,r_0). \label{approx5b}
 \eea
 We find that moments of order larger than $10$ are suppressed
 in these models of neutrino emission.
 The precise number of moments required in study depends on the model of
 emission angle distribution of neutrinos on neutrino sphere.

 The idea that the series of moment equations can be truncated to
 a small set of equations is further supported by examining the
 geometric scaling property of higher moments. For example,
 according to geometric considerations
 $\rho_2$ and $\rho_3$ are suppressed by factors about $1/3^2$
 and about $1/3^4$ compared to $\rho_1$ at $r=3 r_0$ ($\sim 30$km)
 if assuming $\rho_k \sim \rho_0$ on the neutrino sphere.
 A numerical study presented in section \ref{sec8} shows that
 $P_5$ approximation is already quite good in the model described by Eq.
 (\ref{approx4}). More numerical analysis on the truncation of moment
 equations will be presented in other publications.
 We expect that a $P_{10}$ approximation
 is probably enough to describe many phenomena in the transport and
 flavor transformation of supernova neutrinos.

 The numerical analysis is substantially simplified
 when using Eq. (\ref{LVEq5b}) in $P_N$ approximation when
 $N$ is at most as large as around ten.
 In $P_N$ approximation the required computational power is
 proportional to numbers of equations:$2 (N+1) N_e$ where $N_e$ is the number
 of energy bins. In contrast, in the multi-angle simulation the
 number of equations in evolution is $6 N_e N_\theta$ where
 $N_\theta$ is the number of angle bins and the factor $6$ is the
 number of all types of neutrinos and anti-neutrinos. It is a huge amount of
 equations by noticing that $N_e, N_\theta \geq 500$ for $L_\nu=10^{51}$ erg$/$s
 ~\cite{dfcq1}. More angle bins are required for larger neutrino
 luminosity in multi-angle simulation. The approach using moment equations reduces
 the computational power by about two orders of magnitude when $N$ is
 around ten.

 \section{Numerical result}\label{sec8}
 In this section we show some numerical analysis on the non-linear flavor
 transformation of neutrinos. For simplicity we
 neglect matter effect in our analysis.

 The truncated set of Equations (\ref{LVEq5c}) should be carefully
 treated in numerical analysis. Three terms
 in the right hand side of Eq. (\ref{LVEq5c}) have different
 physics. The first term is caused by the diffusion of neutrinos
 and gives correction to the scaling law of moments. For example,
 $Q^1_0$ modifies the scaling of the total flux. The second term
 leaves $Tr[\rho^{\prime}_k]$, $Tr[(\rho^{\prime}_k)^2]$ and
 $Tr[(\rho^{\prime}_k)^3]$ etc invariant. It is responsible for
 the unitary flavor evolution of moments. The third term does not
 change $Tr[\rho^{\prime}_k]$ but modifies $Tr[(\rho^{\prime}_k)^2]$,
 $Tr[(\rho^{\prime}_k)^3]$ etc. This term gives non-unitary
 evolution of neutrino flavor.
 The appearance of the non-unitary term in moment equations
 makes it hard to do numerical computations.

 In the present analysis we simplify the numerical computation
 and search a self-consistent solution to the moment equations (\ref{LVEq5c}).
 We assume that the evolution
 is dominated by the first and the second term in the right hand
 side of Eq. (\ref{LVEq5c}). We search for solution of the
 following form
 \bea
 \rho^\prime_k=\rho^{\prime0}_k+\Delta \rho^\prime_k, \label{solscheme1}
 \eea
 In this approximation $Q^{1,2}_k$ and
 $H_{A,C}$ are all expressed using $\rho^{\prime0}_k$:
 \bea
 Q^{1,2}_k=Q^{1,2}_k(\rho^{\prime0}_k),~~
 H_{A,C}=H_{A,C}(\rho^{\prime0}_k).
 \label{solscheme3}
 \eea
 $\rho^{\prime0}_k$ satisfies
 \bea
 \frac{d \rho^{\prime0}_k}{d r}&& =- r_0^{-1} Q^1_k(\rho^{\prime0}_k)
 -i[H_A,\rho^{\prime0}_k] \label{solscheme2}
 \eea
 For a small step we get
 \bea
 \rho^{\prime0}_k(r+\Delta r)=-\frac{\Delta r}{r_0} Q^1_k(\rho^{\prime0}_k)
 +e^{-i H_A \Delta r} \rho^{\prime0}_k(r) e^{i H_A \Delta r}
 \label{solscheme2b}
 \eea
 $\Delta \rho^\prime_k$ is considered as a small correction and its
 initial condition is set to zero. It is sourced by $\rho^{\prime0}_k$:
 \bea
 \frac{d \Delta \rho^\prime_k}{d r}&& =-i [H_A, \Delta \rho^\prime_k]
 -i [H_C,Q^2_k(\rho^{\prime0}_k)]. \label{solscheme4}
 \eea
 Eq. (\ref{solscheme4}) can be re-written as
 \bea
 \frac{d \Delta \rho^{''}_k}{d r}&& = -i [H'_C,Q^2_k(\rho^{''0}_k)],
 \label{solscheme5}
 \eea
 where
 \bea
 \Delta \rho^{''}_k=M^\dagger_A \Delta \rho'_k M_A,~~
 \rho^{''0}_k=M^\dagger_A \rho^{'0}_k M_A,~~
  H'_C= M^\dagger_A H_C M_A,~~\label{solscheme6}
 \eea
 $M_A=M_A(\rho^{\prime0}_k)$ is the evolution matrix given by
 Hamiltonian $H_A(\rho^{\prime0}_k)$:
 \bea
 \frac{d }{dr }M_A=-i H_A M_A. \label{solscheme7}
 \eea
 Initial condition $M_A=1$ is set. For a small step
 Eq. (\ref{solscheme5}) gives
 \bea
 \Delta \rho^{''}_k(r+\Delta r)-\Delta \rho^{''}_k(r)
 =e^{-i H'_C \Delta r} Q^2_k e^{i H'_C \Delta r} -Q_k^2
 \label{solscheme8}
 \eea
 We expect $\Delta \rho'_k$
 receive small contributions from $\rho^{'0}_k$ and its effect
 on the final result is negligible. Including $\Delta \rho'_k$
 in $Q^{1,2}_k$ and $H_{A,C}$ introduces even smaller corrections.
 In numerical analysis the consistency of this
 approximation should be checked. That is, including $\Delta \rho'_k$
 should not give much modification to $\rho'_k$.

     \begin{figure}
\begin{center}
\includegraphics[height=6.cm,width=10.8cm]{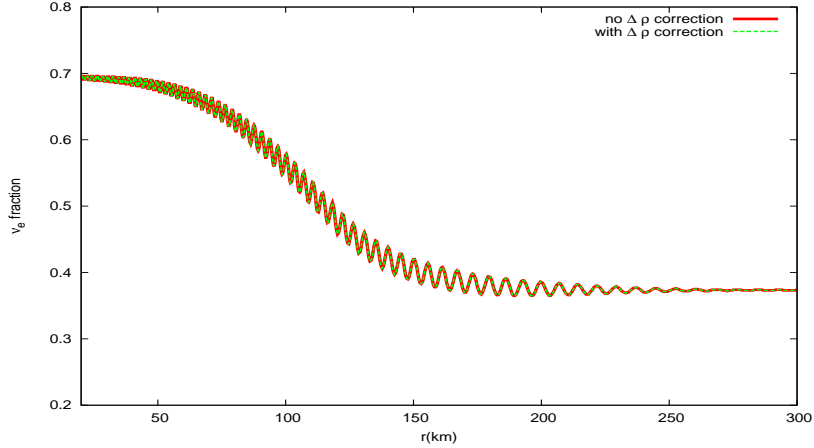}
\end{center}
 \vskip 0.0cm
  \caption{\small (color online) Fraction of $\nu_e$,
  $n_{\nu_e}/(n_{\nu_e}+n_{\nu_x})$, versus radius $r$ in two neutrino system.
  Two lines are for results with and without
   $\Delta \rho'_k$ correction separately.
  $|\Delta m^2_{31}|=3 \times 10^{-3}$eV$^2$, $\sin^2 2\theta_{13}=0.01$.
   }
 \label{fig3}
\end{figure}

 This approximation is supported by the result of numerical computation.
 A result of numerical computation using this approximation is shown
 in Fig. \ref{fig3}. It is shown for two neutrino system of $(\nu_e, \nu_x)$
 with inverted mass hierarchy and for emission angle distribution
 described by Eq. (\ref{approx4}).
 We choose $L_{\nu_e}=L_{{\bar \nu}_e}=L_{\nu_x}=L_{{\bar \nu}_x}
  = 3.\times 10^{51}$ erg$/$s.
  The initial energy spectrum of neutrino is given by the Fermi-Dirac
  distribution
  \bea
  f_\nu(E)=\frac{1}{F_2~T_\nu}
  \frac{x^2}{e^{x-\mu_\nu}+1}, \label{dist}
  \eea
  where $x=E/T_\nu$ and $F_2$ is the normalization factor.
  Parameters of four types of neutrinos are chosen as:
  $T_{\nu_e}=2.76$ MeV, $T_{{\bar \nu}_e}=4.01$ MeV,
  $T_{\nu_x}=T_{{\bar \nu}_x}=6.26$ MeV.
  $\mu_{\nu_e}=\mu_{{\bar \nu}_e}=\mu_{\nu_x}=\mu_{{\bar \nu}_x}=3.$
  In Fig. \ref{fig3} one can see clearly the synchronized oscillation
  and the transition to bipolar oscillation.

  We show both result with $\Delta \rho'_k$ corrections and the result
  without $\Delta \rho'_k$ corrections. As can be seen these two results
  agree very well. $\Delta \rho'_k$ corrections give negligible
  corrections to the neutrino evolution. This can be understood by
  noticing that in the small $r$ region it is dominated by synchronized
  oscillation where all $\rho_{\vec p}$ point to the same direction
  in flavor space and they commute with the effective Hamiltonian $H_C$.
  Hence effect of the third term in Eq. (\ref{LVEq5c}) does not change
  synchronized oscillation and is not important in small r region.
  In the large $r$ region the magnitude of $Q^2_k$ is suppressed by $r_0^2/r^2$ and is
  again not important. This result shows that in this approximation
  we can neglect the third term in Eq. (\ref{LVEq5c}) and use
  the following equation in our analysis
  \bea
 \frac{d \rho^\prime_k}{d r}&& =- r_0^{-1} Q^1_k-i[H_A,\rho^\prime_k].
 \label{LVEq5d}
 \eea

\begin{figure}
\begin{center}
\includegraphics[height=6.cm,width=10.8cm]{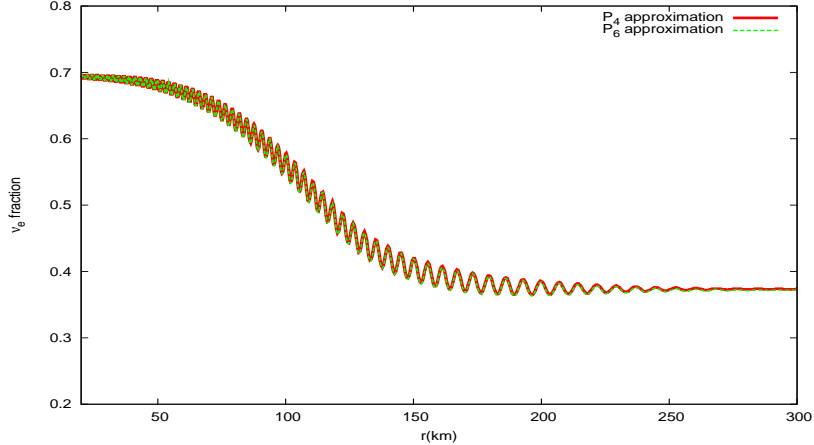}
\end{center}
 \vskip 0.0cm
  \caption{\small (color online) Fraction of $\nu_e$,
  $n_{\nu_e}/(n_{\nu_e}+n_{\nu_x})$, versus radius $r$ in two neutrino system.
   Two lines are for $P_4$ and $P_6$ approximations
   separately. Parameters of neutrinos are the same as in Fig. \ref{fig3}.
   }
 \label{fig4}
\end{figure}

  Result in Fig. \ref{fig3} is obtained
  using $P_4$ approximation. In Fig. \ref{fig4} we compare the result of
  $P_4$ approximation with the result of $P_6$ approximation.
  One can see that results of these two approximations agree quite well.
  This result of numerical analysis shows that truncated moment equations
  indeed have very good convergence property.
  More detailed numerical analysis will be presented in other publications.

 \section{Equations in two flavor system} \label{sec9}
 In two flavor system of neutrinos the density matrix can be
 expressed in terms of the unit matrix and the Pauli matrices.
 It's much easier to do analytic study in two flavor system. In this
 section we re-write the moment equations for two flavor system.

 For two flavors of neutrinos we write
 \bea
&& H_0=\frac{1}{2}(\omega_0+\omega_E {\vec B}\cdot {\vec \sigma}),
 ~~L=\frac{1}{2} (L^0+{\vec L}\cdot \sigma), \label{2flavor1} \\
&& E^2 ~\rho^\prime_k(E)=\frac{1}{2}(P_k^0+{\vec P}_k
 \cdot {\vec \sigma}),
 ~~E^2 ~{\bar \rho}^\prime_k(E)=\frac{1}{2}(P_k^{\prime 0}
 +{\vec P}^\prime_k \cdot {\vec \sigma}). \label{2flavor3}
 \eea
 where $E$ is the energy of neutrino and
 ${\vec \sigma}=(\sigma^1,\sigma^2,\sigma^3)$ are Pauli matrices.

  Equations are obtained when using commutation relation of the
  Pauli matrices. Neglecting moments of $k>1$ and the third term in
  the right hand side of Eq. (\ref{LVEq5c}) we get in the mass base of neutrino
 \bea
 \frac{d {\vec P}_0}{d r} &&= -2 r_0^{-1} z^{-2}
 {\vec P}_1 +[\omega_E{\vec B} +\sqrt{2} G_F ({\vec L}+ z^{-4}
 {\vec N}_1)]\times {\vec P}_0, \label{pendulum3} \\
 \frac{d {\vec P}^\prime_0}{d r} && =-2 r_0^{-1} z^{-2}
 {\vec P}^\prime_1 +[-\omega_E{\vec B} +\sqrt{2} G_F ({\vec L}+ r^{-4}
 {\vec N}_1)]\times {\vec P}^\prime_0,
 \label{pendulum4} \\
 \frac{d {\vec P}_1}{d r} &&=(\omega_E {\vec B}+\sqrt{2} G_F {\vec L}+
 \sqrt{2} G_F z^{-4} {\vec N}_1) \times {\vec P}_1, \label{pendulum1} \\
 \frac{d {\vec P}^\prime_1}{d r} &&=(-\omega_E {\vec B}+\sqrt{2} G_F {\vec L}
 +\sqrt{2} G_F z^{-4} {\vec N}_1) \times {\vec P}^\prime_1. \label{pendulum2}
 \eea
 where $\omega_E=\Delta m^2/2E$ and
 \bea
 {\vec N}_1=\int \frac{d E}{(2\pi)^3} ({\vec P}_1-{\vec P}^\prime_1).
 \eea
 Eqs. (\ref{pendulum3}), (\ref{pendulum4}), (\ref{pendulum1}) and
 (\ref{pendulum2}) form a closed set of equations.

 Eqs. (\ref{pendulum3}), (\ref{pendulum4}), (\ref{pendulum1}) and
 (\ref{pendulum2}) are equations we obtained for describing the pendulum
 motion of neutrinos in flavor space.
 Eqs. (\ref{pendulum1}) and (\ref{pendulum2}) are the equations
 widely used in two flavor analysis in literature~\cite{hrsw}.
 An improvement of Eqs (\ref{pendulum1}) and (\ref{pendulum2})
 is that the dependence on the radius $r$ is explicitly
 derived. Correction to the scaling law of the Hamiltonian
 $z^{-4}{\vec N}$ can be included when higher moments are included.
 This is another virtue using moment equations.

 We note that an important piece of this set
 of equations, Eqs.(\ref{pendulum3}) and (\ref{pendulum4}),
 are missed in the literature. Actually what have been missed
 are more relevant to physical observables. Neutrino densities
 are described by $P^0_0$($P^{'0}_0$) and $P^3_0$($P^{'3}_0$).
 The dynamics of non-linear flavor transformation is controlled by ${\vec P}_1$ and
 ${\vec P}'_1$ which are not real observables. Predictions of these two new
 equations on the flavor transformation should be explored.

 It should  be reminded that Eqs. (\ref{pendulum1}),(\ref{pendulum2}), (\ref{pendulum3})
 and (\ref{pendulum4}) may not be reliably used in precise numerical
 analysis since moments of order $k>1$ have all been neglected when
 deriving them. They can be used in qualitative analysis. We note
 that including the third term in the right hand side of Eq. (\ref{LVEq5c})
 leads to extra term in pendulum equation of neutrinos.

 \section{Conclusion}\label{sec10}

 In summary we derive a series of moment equations describing the transport
 and flavor transformation of neutrinos above neutrino sphere
 in core-collapse supernova. We examine the system of neutrinos in supernova
 using spherical coordinate and find a particular set of moments
 of the density matrix of neutrino. We expand the Liouville
 equation of neutrinos using this series of moments and obtain
 the moment equations. We work with assumptions that the system
 is spherically symmetric and stationary in time scale around
 $10^{-3}$s. We arrive at a truncated set of moment equations
 (\ref{LVEq5c}). After a further approximation we get (\ref{LVEq5d}).
 Rates of neutrino flavor transformation can be solved as functions of radius
 $r$ using this
 set of truncated moment equations. These equations will be very helpful
 to future researches on supernova neutrinos~\cite{kyabs}.

 We study geometric scaling properties of these moments in the
 expanding system of supernova and find that they have very good
 property of convergence. Higher moments $\rho_k$($k>1$) converge to zero
 much faster than lower moments $\rho_{0,1}$. Furthermore,
 we check initial conditions of moments and find that magnitudes of
 higher moments are naturally smaller than $\rho_{0,1}$ on the neutrino
 sphere. Based on these observations we argue that the infinite series of
 moment equations can be safely truncated to be a small set of
 equations with $\rho_{k}(k\lsim 10)$. Numerical analysis shows that
 the results in $P_N$ approximation indeed converge for $N\lsim 10$.
 In Fig. \ref{fig3} and Fig.
 \ref{fig4} we show some numerical analysis and find patterns of synchronized
 oscillation and bipolar oscillation.

 The truncated set of moment equations have a number of good properties.
 First, the truncated set of equations tremendously simplify
 numerical analysis of the problem. Previous numerical works
 use many angle bins and energy bins to simulate the neutrino
 evolution. In multi-angle simulation evolutions of
 more than $500$ trajectories have to be followed. Together with
 more than $500$ energy bins and an extra factor $6$ ( number of
 total types of neutrinos and anti-neutrinos), more than a million
 equations have to be solved
 in this kind of simulation. It is ultra-complicated.
 The approach presented in this article uses a small set of
 truncated moments rather than many angle bins
 and is much simpler than the multi-angle simulation.
 Using a small set of truncated moment equations the required computational power
 in numerical analysis is reduced by two orders of magnitude
 compared to that in multi-angle simulation.

 Second, the moment equations systematically take into
 account the effect of the angular distribution of neutrino emission.
 As noted in Introduction, angular distribution of neutrino
 emission is essential for the effect of neutrino self-interaction
 to play an important in neutrino flavor transformation.
 The approach presented in the present article introduces moments of
 density matrix of neutrinos to describe the angular distribution
 of neutrinos. The equation of zeroth moment naturally
 includes the effect of higher moments
 on the evolution of neutrino intensity. The equations of higher
 moments systematically take into account the evolution of angular
 distribution of neutrino. In particular, modification to the
 scaling of the strength of the effective Hamiltonian can be included when
 including moments $\rho_k$ with $k \geq 2$. The dependence of this
 modification on the emission angle distribution of neutrinos can be
 systematically studied. In contrast, previous researches use fixed
 scaling law (basically $r^{-4}$ law) for the strength of the effective
 Hamiltonian. This set of moment equations provides
 a powerful tool towards a complete understanding to the effect of
 the emission angle distribution in flavor transformation of
 supernova neutrinos. More analysis on the effect of angular
 distribution in neutrino oscillation will be presented in other
 publications.

 We also consider two flavor system of neutrinos. Using the
 truncated set of equations for zeroth and first moments
 we derive equations describing the pendulum motion of neutrinos
 in flavor space. In addition to equations given in literature,
 we find new equations which are important and are
 more relevant to physical observable.

 There are still many problems. More detailed works should
 be done to check carefully the evolution of higher moments
 and their effect in convergence of moment equations.
 More numerical works should be done to analyze in detail
 the effect of higher moments on phenomena such as synchronized
 oscillation and bipolar oscillation, spectral split and
 adiabaticity of neutrino evolution, etc.
 One important remaining problem is how to extend formulation to the case without
 the assumption of spherical symmetry.  Answer to this problem will
 enable us to understand the impact of anisotropic disturbance of ordinary
 matter on neutrino flavor transformation. To answer this question we
 need to examine Eq. (\ref{LVEq2a}) in more detail. Subtle problems
 in quantum evolution of moment equations should
 be investigated before we can answer this question.
 These problems will be explored in future works.
 \\

 Acknowledgement:
 I wish to thank Y. Z. Qian, G. Raffelt, A. Yu. Smirnov for
 discussions on neutrino flavor conversion in supernova.
 \footnote{Note added: After submission of this article I know
 from A. Mirizzi that an earlier attempt to do moment expansion
 of the equation of supernova neutrinos is made in Ref. \cite{RS}.}

\section*{Appendix}

 Neglecting the gravitational potential, the invariant distance
 in spherical coordinate $x^i=(r,\theta,\varphi)$ is given by
 \bea
 ds^2=dt^2-g_{ij} dx^i dx^j, \label{invdistance}
 \eea
 where $g_{ij}$ is the metric
 \bea
 g_{ij}=diag\{1,r^2,r^2\sin^2\theta\}. \label{metric}
 \eea

 The Christoffel symbol $\Gamma^{i}_{jk}$ is computed using
 the metric $g_{ij}$:
 \bea
 \Gamma^{i}_{jk}= \frac{1}{2} g^{il} (\frac{\partial g_{lk}}{\partial x^j}
 +\frac{\partial g_{jl}}{\partial x^k}-\frac{\partial g_{jk}}{\partial x^l}).
 \eea
 We get
 \bea
 && \Gamma^r_{\theta\theta}=-r,~\Gamma^r_{\varphi\varphi}=-r \sin^2\theta,
  ~\Gamma^{\theta}_{r\theta}=\Gamma^{\theta}_{\theta r}=r^{-1}, \\
  && \Gamma^\theta_{\varphi \varphi}=-\sin\theta \cos\theta,
 ~\Gamma^{\varphi}_{r \varphi}=\Gamma^\varphi_{\varphi r}=r^{-1},
 ~\Gamma^\varphi_{\theta \varphi}=\Gamma^\varphi_{\varphi
 \theta}=ctg\theta.
 \eea
 All other $\Gamma$'s are zero.

 Using Eq. (\ref{GeodEq}) we find
 \bea
 && \frac{dp^r}{dt}=\frac{1}{r}\frac{r^2 (p^\theta)^2+r^2 \sin^2\theta
 (p^\varphi)^2 }{ p^0}, \label{GeodEq2}\\
 && \frac{dp^\theta}{dt}=-\frac{2}{r}\frac{p^r p^\theta}{p^0}+
 \sin\theta \cos\theta \frac{(p^\varphi)^2}{p^0}, \label{GeodEq3}\\
 && \frac{dp^\varphi}{dt}=-\frac{2}{r}\frac{p^r p^\varphi}{p^0}
 -2~ctg\theta ~\frac{p^\theta p^\varphi}{p^0}, \label{GeodEq4}
 \eea
 It's easy to verify that
 \bea
 \frac{d}{dt}[(p^r)^2+r^2 (p^\theta)^2+r^2
 \sin^2\theta (p^\varphi)^2]=0.
 \eea
 This is just the statement that $|{\vec p}|=\sqrt{g_{ij} p^i p^j}
 =\sqrt{(p^r)^2+r^2 (p^\theta)^2+r^2 \sin^2\theta (p^\varphi)^2}$,
 the magnitude of the momentum, is a constant. It equals to $p^0$
 for neutrino with tiny masses.

 $p^r$, $\sqrt{g_{\theta\theta}} p^{\theta}=r p^{\theta}$ and
 $\sqrt{g_{\varphi\varphi}} p^{\varphi}=r \sin\theta p^{\varphi}$
 are the momenta projected to ${\hat r}$, ${\hat \theta}$ and ${\hat \varphi}$
 directions separately. Hence we can write
 \bea
 p^r=\cos\theta_p ~p^0,~p^{\theta}=\frac{1}{r}\sin\theta_p \cos\varphi_p ~p^0,
 ~p^\varphi=\frac{1}{r \sin\theta} \sin\theta_p \sin\varphi_p ~p^0.
 \label{momentum}
 \eea
 Using Eq. (\ref{GeodEq2}) and (\ref{GeodEq3}) we get
 \bea
 \frac{d \cos\theta_p}{d t}= \frac{1}{r}(1-\cos^2\theta_p),
 \label{GeodEq5} \\
 \frac{d \varphi_p}{d t}=-\frac{1}{r} ctg\theta ~\sin\theta_p
 ~\sin\varphi_p. \label{GeodEq6}
 \eea
 The consequence of Eq. $(\ref{GeodEq5})$ is that $\cos\theta_p$
 approaches to $1$. This is consistent with the physical
 picture that as neutrinos go out of the supernova their motion
 gets closer to the radial direction.

\end{document}